\documentclass[pre, aps,twocolumn,floatfix,superscriptaddress]{revtex4-1}
\usepackage{amsmath,amssymb,graphicx,cases}
\usepackage{epsfig}
\usepackage{textcomp}
\usepackage{epstopdf}
\usepackage{float}
\usepackage{braket}
\usepackage[normalem]{ulem}

\usepackage[normalem]{ulem}
\newcommand{\stkout}[1]{\ifmmode\text{\sout{\ensuremath{#1}}}\else\sout{#1}\fi}

\usepackage{fixmath}

\usepackage{color}
\definecolor{Blue}{rgb}{0.00, 0.00, 1.00}
\definecolor{Red}{rgb}{1.00, 0.00, 0.00}
\definecolor{Green}{rgb}{0.00, 0.60, 0.00}

\newcommand{\nn}{\nonumber}
\newcommand{\be}{\begin{equation}}
\newcommand{\ee}{\end{equation}}
\newcommand{\bea}{\begin{eqnarray}}
\newcommand{\eea}{\end{eqnarray}}

\usepackage[colorlinks=true, urlcolor=blue, anchorcolor=blue, citecolor=blue,filecolor=blue,linkcolor=blue,menucolor=blue]{hyperref}

\begin{document}
\title{Confined run and tumble particles with non-Markovian tumbling
  statistics}

\author{Oded Farago}
\email{ofarago@bgu.ac.il}

\affiliation{Department of Biomedical Engineering, Ben-Gurion
  University of the Negev, Marcus Family Campus Be'er Sheva, 8410501,
  Israel}

\author{Naftali R. Smith}
\email{naftalismith@gmail.com}

\affiliation{Department of Environmental Physics,
  Blaustein Institutes for Desert Research, Ben-Gurion University of
  the Negev, Sede Boqer Campus, 8499000, Israel}


\begin{abstract}

Confined active particles constitute simple, yet realistic, examples
of systems that converge into a non-equilibrium steady state.  We
investigate a run-and-tumble particle in one spatial dimension,
trapped by an external potential, with a given distribution $g(t)$ of
waiting times between tumbling events whose mean value is equal to
$\tau$. Unless $g(t)$ is an exponential distribution (corresponding to
a constant tumbling rate), the process is non-Markovian, which makes
the analysis of the model particularly challenging.  We use an
analytical framework involving effective position-dependent tumbling
rates, to develop a numerical method that yields the full steady-state
distribution (SSD) of the particle's position.  The method is very
efficient and requires modest computing resources, including in the
large-deviations and/or small-$\tau$ regime, where the SSD can be
related to the the large-deviation function, $s(x)$, via the scaling
relation $P_{{\rm st}}(x)\sim e^{-s\left(x\right)/\tau}$.

\end{abstract}

\maketitle

\section{Introduction}



Active particles \cite{Romanczuk,soft,BechingerRev, FodorEtAl15,
  Fodor16, Needleman17, Ramaswamy2017,Marchetti2017,Schweitzer,
  FJC22}, in contrast with their passive Brownian counterparts, use
energy that they pump out of their environment in order to
move. Activity leads to breaking of time-reversal symmetry, and as a
result, even single active particles are out of (thermal)
equilibrium. Many natural examples of active systems may be found in
biology, from molecular motors \cite{BHG06, Toyota11, Stuhrmann12,
  Mizuno07} to cells \cite{Berg2004,Wilhelm08, Cates2012, Ahmed15,
  BeerEtAl20, Breoni22, Nishiguchi23}, to animals \cite{flocking1,
  flocking2, Vicsek,fish}. Moreover, active systems have been
experimentally realized and studied. This has been done by fabricating
artificial particles (such as Janus particles) or robots that propel
themselves \cite{BechingerRev,
  Hagen2014,Takatori2016,Deblais2018,Dauchot2019, gran2, GLA07,
  PBDN21, BCASL22, MolodtsovaEtAl23, PPSK23, Nishiguchi23} leading to
behavior which is similar to that of the natural examples.

 The statistical behavior of active systems may be quite
remarkable indeed, and very different to the equilibrium behavior
exhibited by passive systems, both for active, many-body systems
(``active matter'') \cite{separation1, separation2, separation3,
  cluster1,cluster2,evans, Kardar2015, PBV19, Singh21, ARYL21,
  Cates22, ARYL23, SSB22, ACJ22} and even at the single particle
level \cite{ Wensink2008, Elgeti2009,gran2,  Li2009b,TC08, Kaiser2012,
  Elgeti2013, Hennes2014, Solon2015, Takatori2016, Li2017, Razin2017,
  Dauchot2019, Pototsky2012, Solon2015,Pototsky2012,TC08,
  ABP2019,Dhar_2019,Malakar20,
  Franosch2016,Das2018,Caprini2019,Sevilla2019,
  Hagen2014,Takatori2016,Deblais2018,Dauchot2019, SBS21,
  MalakarEtAl18,ABP2018, Singh2019, SmithFarago22, SLMS22RTP,
  Smith2023, Frydel24, BM24}. A setting that has attracted much recent
interest is that of a single active particle trapped by an external
potential. Such a particle eventually reaches a nonequilibrium steady
state that differs from the Boltzmann-Gibbs statistics
\cite{Solon2015,Pototsky2012,TC08, ABP2019,Dhar_2019,Malakar20,
  Franosch2016,Das2018,Caprini2019,Sevilla2019,
  Hagen2014,Takatori2016,Deblais2018,Dauchot2019, SBS21,
  SmithFarago22, SLMS22RTP, Smith2023, Frydel24}, its first-passage
properties do not follow the Arrhenius law
\cite{MalakarEtAl18,ABP2018, Singh2019}, and it may acquire an
effective drift even if it is placed in a periodic potential
\cite{led20}.

The nonequilibrium nature of active matter makes its analytical study
very challenging. Some relatively simple models admit exact
solutions. One example is the run-and-tumble particle (RTP), which, in
the absence of external forces, moves at a constant speed but changes
its orientation at a constant rate. In one dimension, the RTP model is
exactly solvable with \cite{Kardar2015, Dhar_2019} or without
\cite{wei02,HV_2010,ODA_1988,MADB_2012,MalakarEtAl18,EM_2018,Dhar_2019,SBS20,
  Dean21} an external potential, and in some special cases, these
results may be extended to higher dimensions \cite{SLMS22RTP, ABP2018,
  Malakar20, SBS20, CO21, Frydel22}.
In some less simple settings, active systems are amenable to a
perturbative treatment in various regimes \cite{MB20, SmithFarago22,
  SBS21, Smith2023}.  However, in many cases one must resort to
numerical methods. The challenge posed by active systems becomes
especially pronounced when studying large deviations (i.e., rare
events), which often cannot be studied by brute-force Langevin
dynamics simulations. Large-deviations have become a major theme of
ongoing interest in statistical mechanics \cite{DZ, Hollander,
  Majumdar2007, hugo2009, Derrida11, MajumdarSchehr2017,
  Touchette2018}, and in particular, large deviations in active systems are important because they are often very strongly affected by the
activity of the system, while the effects on the typical fluctuations
may be weaker.

In a vast majority of previous works, the stochastic dynamics of
active particles that were studied were assumed to be Markovian. By
this we mean that the particle's instantaneous position and internal
state (e.g., for an RTP the internal state is its orientation) contain
sufficient information to predict the statistics of its future
dynamics, and it is not necessary to know the full history of the
particle's dynamics  (see footnote~\cite{footnote1}).
 Markovian dynamics are ``memory-free'', and the Markovian
assumption facilitates the analysis considerably, but it also
restricts the class of models that may be analyzed. It is also not so
obvious if this assumption indeed holds for realistic systems, such as
living cells, for which one may reasonably expect that some effects of
memory are present,
see Ref.~\cite{Detcheverry17} and references therein.

In the present work, we consider an RTP with a given distribution
$g(t)$ of waiting times between tumbling events, with mean waiting
time $\tau$. The particular case where $g(t)$ is an exponential
distribution corresponds to a constant tumbling rate, recovering the
standard (Markovian) RTP model.
For any non-exponential $g(t)$, the model becomes non Markovian, and
very challenging to study.  Our model may be thought of as an
extension of the continuous-time random walk model, whose
large-deviations properties have attracted much interest recently
\cite{WBB20, SB23}.  In order to study the model, we develop a
numerical method which allows to compute the nonequilibrium steady
state distribution (SSD) of the particle,
including in
the regime of large deviations and/or small
$\tau$.  Many of the existing large-deviations numerical methods are
based on generating realizations of the stochastic process under study
that are biased toward the rare event of interest \cite{Hartmann2002,
  TL09, HG11, touchette2011, Claussen2015, Hartmann2018, GV19,
  Hartmann2019LIS, Hartmann2019,Hartmann2020,HartmannMS2021, CBG23,
  SmithChaos22, SmithFarago22, NickelsenTouchette22}. In contrast, our
method circumvents the need to generate realizations of the process,
while relying on some theoretical insights that we derive. It should
thus be viewed as a combined theoretical and numerical framework.

The remainder of the paper is organized as follows.  In section
\ref{sec:theory} we define the non-Markovian RTP model precisely, and
introduce the key theoretical concepts for studying the nonequilibrium
steady state that is reached when this particle is placed in a
trapping potential.  In section \ref{sec:numerical} we use these
concepts to describe the numerical method by which the steady state
can be computed.  In section \ref{sec:results} we present results that
we obtained using this method for three particular waiting time
distributions $g(t)$: Exponential (which serves as a benchmark to
verify that our results agree with the known, exact solution),
semi-Gaussian, and half-t.  In section \ref{sec:disc} we summarize and
discuss our main findings.

\section{Model definition and theoretical framework}
\label{sec:theory}

We consider an overdamped particle moving in a one-dimensional (1D)
potential field $U(x)$. For simplicity we assume that $U(x)$ is mirror
symmetric $U(x) = U(-x)$, and has a unique minimum at $x=0$. The
particle is subject to an active telegraphic (dichotomous) noise that
switches between two values $\sigma(t)=\pm \sigma_0$. This is known as
the 1D run-and-tumble particle (RTP) model. We further assume that the
running times [i.e., the times between consecutive switching events of
  the sign of $\sigma(t)$] are drawn from a general distribution
$g(t)$ [$\int_0^{\infty} g(t)dt=1$]. We denote the  mean
running time by $\tau = \int_0^\infty t g(t) dt$. The most thoroughly studied case is when
$g(t)=\exp(-t/\tau)/\tau$, for which $\sigma(t)$ 
describes tumbling (switching) at a constant (time-independent)
rate $\gamma=1/\tau$. For any
other distribution function, $g(t)$, the process is non-Markovian.

The Langevin equation describing the dynamics of the particle is
\begin{equation}
  \mu \dot{x}-F(x)-\sigma(t)=0,
\label{eq:langevin}
\end{equation}
where $\mu$ is the friction coefficient, and
$F(x)=-U^{\prime}(x)$. Since the particle has no inertia, it is
confined to the interval $-X_0<x<X_0$, where $|F(X_0)|=\sigma_0$
[assuming that such $X_0$ exists, and that $-X_0<x(t=0)<X_0$]. Its
instantaneous direction of motion is that of $\sigma(t)$, and for a
given orientation, the speed is an injective function of $x$ given by
\begin{equation}
  \mu v_{\pm}(x)=\sigma_0\pm F(x),
  \label{eq:vpm}
\end{equation}
where the subscripts $\pm$ denote the speeds when the particle moves in the
positive and negative directions, respectively.

Here, we introduce a theoretical-numerical framework for calculating
the SSD of the particle, $P_{\rm st}(x)$. Our approach is not based on
straightforward Langevin dynamics simulations, which tend to be
inefficient in the large-deviation regime $|x|\lesssim X_0$,
especially when $\tau$ is small. Instead, we begin the analysis by
considering the probability flux which is constant at steady-state,
$J_{\rm st}(x)=J$. Since the particle is confined to a finite
interval, we have that $J_{\rm st}=0$. In what follows, we consider
the steady-state statistics of the particle and, therefore, drop the
subscript ``st'' for brevity.

\subsection{Steady-state currents}

The steady-state flux is the difference between the currents of
particles moving in the positive and negative directions,
$J(x)=I_+(x)-I_-(x)$. We denote by $P_{+}(x)$ and $P_{-}(x)$ the
distributions corresponding, in steady-state, to right- and
left-moving particles, respectively. These are related to the SSD
$P(x)$ via
\begin{equation}
  P(x)=P_+(x)+P_-(x),
\label{eq:pplusminus}
\end{equation}
and because of the symmetry of the problem, both of them are
normalized to $\int_{-\infty}^{\infty}P_{\pm}\left(x\right)dx=1/2$
(rather than unity), and satisfy $P_+(x)=P_-(-x)$. The associated
currents are given by $I_{\pm}(x)=P_{\pm}(x)v_{\pm}(x)$.  Since $J=0$,
we have that $I_+(x)=I_-(x)$, and so we adopt the same notation $I(x)$
for both, i.e.,
\begin{equation} I(x)=P_{+}(x)v_{+}(x)=P_{-}(x)v_{-}(x).
\label{eq:currents}
\end{equation}

We now arrive at a key part of the derivation, which is the
introduction of the space-dependent effective switching rates,
$\gamma_{\pm}(x)$. {\em These rates express the probability, per unit
  time, of a particle to tumble (i.e., flip its orientation) when
  traveling in the vicinity of $x$}. We denote by $i_{\pm}^t(x)$, the
associated tumbling current densities (per unit length per unit time)
\begin{equation}
  i_{\pm}^t(x)=\gamma_{\pm}(x)P_{\pm}(x),
  \label{eq:probstop}
\end{equation}
which from Eq.~(\ref{eq:currents}) can be also written as
\begin{equation}
  i_{\pm}^t(x)=I(x)\frac{\gamma_{\pm}(x)}{v_{\pm}(x)}.
  \label{eq:probstop2}
\end{equation}
Due to the occasional switches in the direction of motion, the current
changes, over a small distance $dx$, by
$dI(x)=\left[i_{-}^t(x)-i_+^t(x)\right]dx$. Thus, we arrive at the
differential equation
\begin{equation}
  \frac{dI}{dx}=I(x)\left[\frac{\gamma_-(x)}{v_-(x)}
  -\frac{\gamma_+(x)}{v_+(x)}\right],
  \label{eq:current1}
\end{equation}
with the solution
\begin{equation}
  I(x)=I(x=0)\exp\left\{\int_0^x\left(\frac{\gamma_-(y)}{v_-(y)}
  -\frac{\gamma_+(y)}{v_+(y)}\right)dy\right\}.
  \label{eq:current2}
\end{equation}
From Eqs.~(\ref{eq:pplusminus}) and
(\ref{eq:currents}), we have that
\begin{equation}
  P(x)=I(x)\left(\frac{1}{v_+(x)}+\frac{1}{v_-(x)}\right).
  \label{eq:current3}
\end{equation}
Furthermore, because $F(0)=0$, we have from Eq.~(\ref{eq:vpm}) that
$v_{\pm}(0)=\sigma_0/\mu$, and we thus arrive at the result that
\begin{eqnarray}
  \frac{P(x)}{P(x=0)}&=&\frac{\sigma_0^2}{\sigma_0^2-F^2(x)}\times
    \label{eq:ssd1} \\[1mm]
  &\exp&
  \left\{\mu\int_0^x\left(\frac{\gamma_-(y)}{\sigma_0-F(y)}
  -\frac{\gamma_+(y)}{\sigma_0+F(y)}\right)dy\right\}\nonumber. 
\end{eqnarray}

\subsection{Relating $\gamma_{\pm}(x)$ to $g(t)$} 

A derivation similar to the one presented in the previous subsection
was introduced in Ref.~\cite{Monthus2021}. However, in that paper the
rates $\gamma_{\pm}(x)$ were simply
taken to be given functions, i.e., the model studied there was a
Markovian RTP with space-dependent tumbling rates (and velocities). In
contrast, in our non-Markovian model, $\gamma_{\pm}(x)$ are a-priori
unknown, but we can relate them to the function $g(t)$, which
characterizes the statistics of switching times.

When $g(t)=\exp(-t/\tau)/\tau$, the switching rates are both time- and
space-independent, i.e., $\gamma_{\pm}(x)=\gamma=1/\tau$, and we
recover the well-known result
\cite{q-optics1,q-optics2,q-optics3,q-optics4, VBH84, colored,
  Kardar2015, Dhar_2019}
\begin{eqnarray}
  \frac{P(x)}{P(x=0)}&=&\frac{\sigma_0^2}{\sigma_0^2-F^2(x)}\times
    \label{eq:ssd2}  \\
  &\exp&
    \left\{\frac{2\mu}{\tau}\int_0^x\left(\frac{F(y)}
                {\sigma_0^2-F^2(y)}\right)dy\right\}\nonumber.
\end{eqnarray}
To proceed in the general case, we introduce the transition
probability density per unit length, $\Pi_+(a,b)$, that the particle
travels directly (i.e., without switching directions in between),
starting at point $x=a$ and stopping at point $x=b>a$. Similarly,
$\Pi_-(a,b)$ denotes the probability density of a movement interval
that starts at $x=a$ and ends at $x=b<a$. Due to the symmetry of the
problem with respect to $x=0$, it is clear that
$\Pi_+(a,b)=\Pi_-(-a,-b)$. The tumbling current densities
\eqref{eq:probstop} satisfy the following set of coupled equations
\begin{eqnarray}
  i_+^t(x)&=&\int_{-X_0}^{x}i_-^t(y)\Pi_+(y,x)dy,\label{eq:probstop3}\\
  i_-^t(x)&=&\int_{x}^{+X_0}i_+^t(y)\Pi_-(y,x)dy.\label{eq:probstop4}
\end{eqnarray}
These equations are obtained by integrating over the position $y$ of
the last tumbling event before the tumbling at position $x$. In order
to express the transition probability densities, we consider the
traveling time along the interval from from $x=a$ to $x=b$, which is
given by
\begin{equation}
  t_+(a,b)=\int_a^b \frac{dx}{v_+(x)},
  \label{eq:tplusab}
\end{equation}
if $a<b$, and by 
\begin{equation}
  t_-(a,b)=t_+(-a,-b),
  \label{eq:tminusba}
\end{equation}
if $a>b$. Since the particle always moves at the instantaneous
direction of the noise, the probability densities associated with
these traveling times are $g[t_{\pm}(a,b)]$. These probability
densities per unit time are related to the transition densities per
unit length, via 
\begin{equation}
  \Pi_{\pm}(y,x)=g\left[t_{\pm}\left(y,x\right)\right]\frac{\partial
    t_{\pm}}{\partial y}=
  \frac{g\left[t_{\pm}\left(y,x\right)\right]}{v_{\pm}(y)}.
  \label{eq:gtopi}
\end{equation}
From Eqs.~(\ref{eq:probstop3}), (\ref{eq:probstop4}), and
(\ref{eq:gtopi}) we find that
\begin{eqnarray}
  i_+^t(x)&=&\int_{-X_0}^{x}i_-^t(y)\frac{g
  \left[t_+\left(y,x\right)\right]}{v_+(y)}dy,\label{eq:probstop5}\\[1mm]
  i_-^t(x)&=&\int_{x}^{X_0}i_+^t(y)\frac{g
  \left[t_-\left(y,x\right)\right]}{v_-(y)}dy.\label{eq:probstop6}
\end{eqnarray}
Eqs.~(\ref{eq:probstop5}) and (\ref{eq:probstop6}) are a set of
integral equations for the tumbling current densities $i_{\pm}^t(x)$,
and they constitute the main results of the theoretical part of our
analysis. In these equations, the function $g(t)$ is given and so are
$v_{\pm}(x)$, as they are given immediately from the external force
$F(x)$ via Eq.~\eqref{eq:vpm}.  Given the solution $i_{\pm}^t(x)$ to
these equations, one can obtain $I(x)$ by integrating the equation
$dI/dx=i_{-}^t(x)-i_+^t(x)$, yielding the steady-state distribution
$P(x)$ through Eq.~\eqref{eq:current3}.  However, solving
Eqs.~(\ref{eq:probstop5}) and (\ref{eq:probstop6}) for general $g(t)$
and $F(x)$ is a  highly non-trivial task. In the next section,
we present a numerical method for solving these equations.

\section{The Numerical Scheme}
\label{sec:numerical}

\subsection{Computing the tumbling current densities $i_{\pm}^t(x)$}

The set of integral equations
(\ref{eq:probstop5})-(\ref{eq:probstop6}) can be solved
numerically. For this purpose, we discretize the support interval
$(-X_0,+X_0)$ into an even number $N$ of small bins of size $\delta
x=2X_0/N$. Denoting the end points by $x_0=-X_0$ and $x_{N}=+X_0$, we
use the integer variable $i$ to index the bin extending from $x_{i-1}$
to $x_{i}=x_{i-1}+\delta x$.  We then determine the travel time
between any two points $x_i$ and $x_j$ for $i,j=1,2,\ldots,N$. This
can be done either analytically (if possible) or numerically. Because
of the symmetry of the problem, we only consider the travel times in
the positive direction. In the case of a numerical integration, it is
possible to use Eq.~(\ref{eq:tplusab}) to determine the travel time,
$t_+(j,i)$, between $x_j$ and $x_i\geq x_j$, but the vanishing of
$v_+(x)$ close to the edge of the support, $x\lesssim X_0$, imposes
severe restrictions on the bin size, $\delta x$. Therefore, it is
recommended to compute $t_+(j,i)$ by numerically integrating the
equation of motion (\ref{eq:langevin}) with a small integration
time-step $\delta t$. Note that in this process, the Langevin equation
(\ref{eq:langevin}) is not treated as a stochastic, but as a
deterministic equation, with $\sigma(t)=+\sigma_0$ for a particle
moving rightwards.

It is tempting to write the discretized version of
Eq.~\eqref{eq:probstop5} as
\begin{equation}
  i_+^t(x_i)=\sum_{j=1}^{i} i_-^t(x_j)
  \frac{g\left[t_+\left(j,i\right)\right]}{v_+(x_j)}\delta x.
  \label{eq:rec0}
\end{equation}
 However, similarly to the considerations discussed above
  regarding the computation of the travel times, we must keep in mind
  here that Eq.~(\ref{eq:probstop5}) is a re-expression of
Eq.~(\ref{eq:probstop3}) where the transition probability per unit
length, $\Pi_+(y,x)$, is replaced with the transition probability per
unit time, $g[t_+(y,x)]$. The conversion between them is the
origin of the factor $1/v_+(y)$ appearing in the integrand in
Eq.~(\ref{eq:probstop5}) [see Eq.~(\ref{eq:gtopi})], which is the
inverse ratio between the length of an infinitesimal element around
$y$ and the infinitesimal time that the particle spends in that
element. This differential form is valid only if the travel time is
small, which is not the case when particle approaches the edge of the
support $y\lesssim X_0$. This problem can be solved by noticing that
\bea
\label{eq:deltag}
\int_{x_{i-1}}^{x_{i}}\frac{g\left[t_{+}\left(y,x\right)\right]}{v_{+}(y)}dy&=&G\left[t_{+}(j,i)\right]-G\left[t_{+}(j,i-1)\right]\nn\\
&\equiv&\delta G(j,i),
\eea
where $G(t)=\int_0^t g(s)ds$ is the cumulative distribution function
of the traveling time $t$. We identify the function $\delta G(j,i)$ as
the probability of a particle to stop in the $i$-th bin if the
movement forward begins at $x_j\leq x_{i-1}$. Note that for any $j<N$,
$\delta G(j,N)=1-G[t_+(j,N-1)/\tau]$, because the travel time to
$x_N=+X_0$ diverges. With these considerations, the discretized version
of Eq.~(\ref{eq:probstop5}) is given by
\begin{equation}
  i_+^t(x_i)=\sum_{j=1}^{i} i_-^t(x_j)\delta G(j,i+1),
    \label{eq:rec1}
\end{equation}
which holds for $i=1,2,\ldots N-1$ (i.e., excluding the edges
$x_0=-X_0$ and $x_N=+X_0$). Using the symmetry of the problem
$i_+^t(x)=i_-^t(-x)$, we rewrite Eq.~(\ref{eq:rec1}) as
\begin{equation}
  p_+^t(x_i)=\sum_{j=1}^{i} p_+^t(x_{N-j})\delta G(j,i+1).
    \label{eq:recpstop}
\end{equation}
Notice that in the last equation, we replaced the notation
$i_+^t(x_i)$ with $p_+^t(x_i)$. This is done in order to highlight
that $p_+^t(x_i)$ is the tumbling probability, rather than the
tumbling current, of a right-moving particle in the $i$-th bin. In
(\ref{eq:recpstop}), we have a set of $(N-1)$ linear equations which
can be solved by various numerical techniques. We use a simple
iterative method, starting from the initial state, $p_+^t(x_i^n)=p_0$,
and iterating the set of equations until the stationary solution is
obtained, typically after several tens of iterations. Notice that the
numerically derived probability is not properly normalized but, as we
show below, the normalization cancels when we attempt to estimate the
switching rates $\gamma(x_i)$ and the SSD relative to the origin
$P_+(x_i)/P_+(x_{N/2}=0)$.


\subsection{Computing the SSD and the switching rates}

From the solution for the tumbling
probabilities, $p_+^t(x_i)$, we can evaluate the SSD and the switching
rates using the following expressions: Similarly to the logic behind
the derivation of Eq.~(\ref{eq:recpstop}), it is easy to see that the
discretized steady-state distribution is given by
\begin{equation}
  P_+(x_i)=\sum_{j=1}^{i} p_+^t(x_{N-j})\left\{1-G
  \left[t_+\left(j,i\right)\right]\right\}.
  \label{eq:recssd}
\end{equation}
After the evaluation of $P_+(x_i)$, the discrete steady-state
probability should be normalized to half. Here, we skip this step
because we are only interested in the probability relative to the
center of the potential well.

While $p_+^t(x_i)$ measures the distribution of tumbling points, the
ratio $p_+^t(x_i)/P(x_i)$ measures the tumbling probability of a
right-moving particle located in the $i$-th bin to tumble during the
time interval that it travels through the bin. This quantity is
proportional to the switching rate, $\gamma_+(x_i)$. Explicitly, the
switching rate is obtained by dividing this ratio by the time,
$t_+(i-1,i)$, that the particle spends in the bin
\begin{equation}
  \gamma_+(x_i)=\frac{1}{t_+(i-1,i)}\frac{p_+^t(x_i)}{P_+(x_i)}.
  \label{eq:comgamma0}
\end{equation}
This expression is also insensitive to the normalization of $P_+(x_i)$
because, as evident from Eq.~(\ref{eq:recssd}), the SSD normalization
factor is proportional to normalization factor of the tumbling
probabilities. Eq.~(\ref{eq:comgamma0}) suffers from the same problem
encountered in Eq.~(\ref{eq:rec0}), namely the fact that it assumes
that the travel time in the bin is small. This is not the case near
the edge of the support where the travel times become increasingly
large. Since the tumbling ratio $p_+^t(x_i)/P(x_i)\leq 1$, it implies
that Eq.~(\ref{eq:comgamma0}) would yield switching rates that become
vanishingly small. To fix this problem, we compute the effective
switching rate from the following equation
\begin{equation}
  \frac{p_+^t(x_i)}{P_+(x_i)}=1-\exp\left[-\gamma_+(x_i)t_+(i-1,i)\right],
  \label{eq:comgamma1}
\end{equation}
which compares the computed stopping ratio to the corresponding ratio
in the case that the switching rate is constant during the travel time
[i.e., as if the distribution of running times is exponential with
  $\tau=1/\gamma_+(x_i)$]. As can be easily seen,
Eqs.~(\ref{eq:comgamma0}) and (\ref{eq:comgamma1}) coincide in the
limit $\gamma_+(x_i)t_+(i-1,i)\ll 1$.

\section{Computational Results}
\label{sec:results}

In this section we demonstrate how the noise statistics of
  running times influences the SSD and the effective switching
  rates. For this purpose, we consider a particle in a potential
field given by $U(x)=(2/\pi)\{1-\cos[(\pi/2)x]\}$, which means that
\begin{equation}
  F(x)=-\sin[(\pi/2)x].
  \label{eq:force}
\end{equation}
We set the noise amplitude to $\sigma_0=1$ and the friction
coefficient to $\mu=1.75$. For these parameters, the particle is bound
in the interval $|x|<X_0=1$, which is discretized into bins of size
$\delta x=10^{-3}$. The travel times in the present case can be
calculated analytically:
\begin{eqnarray}
    t_+(j,i)&=&\mu\int_{x_j}^{x_i}
    \frac{dx}{1-\sin[(\pi/2)x]}\label{eq:tsin}\\ &=&\frac{2\mu}{\pi}
    \left\{\tan\left[\frac{\pi}{4}
      \left(1+x_i\right)\right]-\tan\left[\frac{\pi}{4}
      \left(1+x_j\right)\right]\right\} \, .\nonumber
\end{eqnarray}
We also measured $t_+(j,i)$ numerically, by integration of Langevin
equation (\ref{eq:langevin}) with time step $\delta t=10^{-6}$. The
numerical integration results were in full agreement with
Eq.~(\ref{eq:tsin}).

 We study three different distribution functions, $g(t)$, of
  running times: (i) exponential, (ii) semi-Gaussian, and
  (iii) half-t. The first example serves as a way to test the accuracy
  of the method. The other two case studies feature, respectively,
  distribution functions decaying faster and (much) slower than the
  exponential distribution with the same average waiting times. As we
will see below, the computational method introduced above is not only
very accurate including in the large-deviation regime, it is also very
efficient and requires very modest computing resources. One only needs
to compute the travel time, $t_+(j,i)$, between any pair of bins
$j\leq i$, which for the simulations presented herein was done on a PC
in less than 24 hours of CPU time. Once the times are computed, they
can be used for the evaluation of the cumulative function
$G[t_+(j,i)]$ of any distribution function $g(t)$, and the iterations
needed to find the solution of the set of equations
(\ref{eq:recpstop}) take only a few seconds.

\subsection{Exponential distribution (Markov process)}
\label{subsec:exp}

We begin by considering the (exactly-solvable) case
$g(t)=\exp(-t/\tau)/\tau$. For this distribution function of travel
times, the solution for the SSD is given by
Eq.~(\ref{eq:ssd2}). Introducing the
function 
\begin{equation}
  s_{\tau}(x)\equiv-\tau\ln\left[\frac{I(x)}{I(x=0)}\right],
  \label{eq:ldfdef}
\end{equation}
we rewrite Eq.~(\ref{eq:ssd2}) as
\begin{eqnarray}
  \frac{s_{\tau}(x)}{\tau}&=&-\ln\left[\frac{\sigma_0^2-F^2(x)}
    {\sigma_0^2}\frac{P(x)}{P(x=0)}\right]
 \nonumber \\
 \ &=& \frac{2\mu}{\tau}\int_0^x\left(\frac{F(y)}
              {\sigma_0^2-F^2(y)}\right)dy \, . \label{eq:ldf1}
\end{eqnarray}  
We thus find that for the exponential $g(t)$, $s_{\tau}(x)$ is
independent of $\tau$, and coincides with the large-deviation-function
(LDF) $s(x)$ that describes the SSD in the limit $\tau \to 0$ via
$P\left(x\right)\sim e^{-s\left(x\right)/\tau}$ (which is a
large-deviations principle) \cite{SmithFarago22, Smith2023}.  For the
force function (\ref{eq:force}), it is given by
\begin{equation}
  s(x)=\frac{4\mu}{\pi}\left[\frac{1-\cos\left(\frac{\pi}{2}x\right)}
    {\cos\left(\frac{\pi}{2}x\right)}\right].
    \label{eq:ldf2}
\end{equation}

\begin{figure}[t]
\includegraphics[width=0.98\linewidth,clip=]{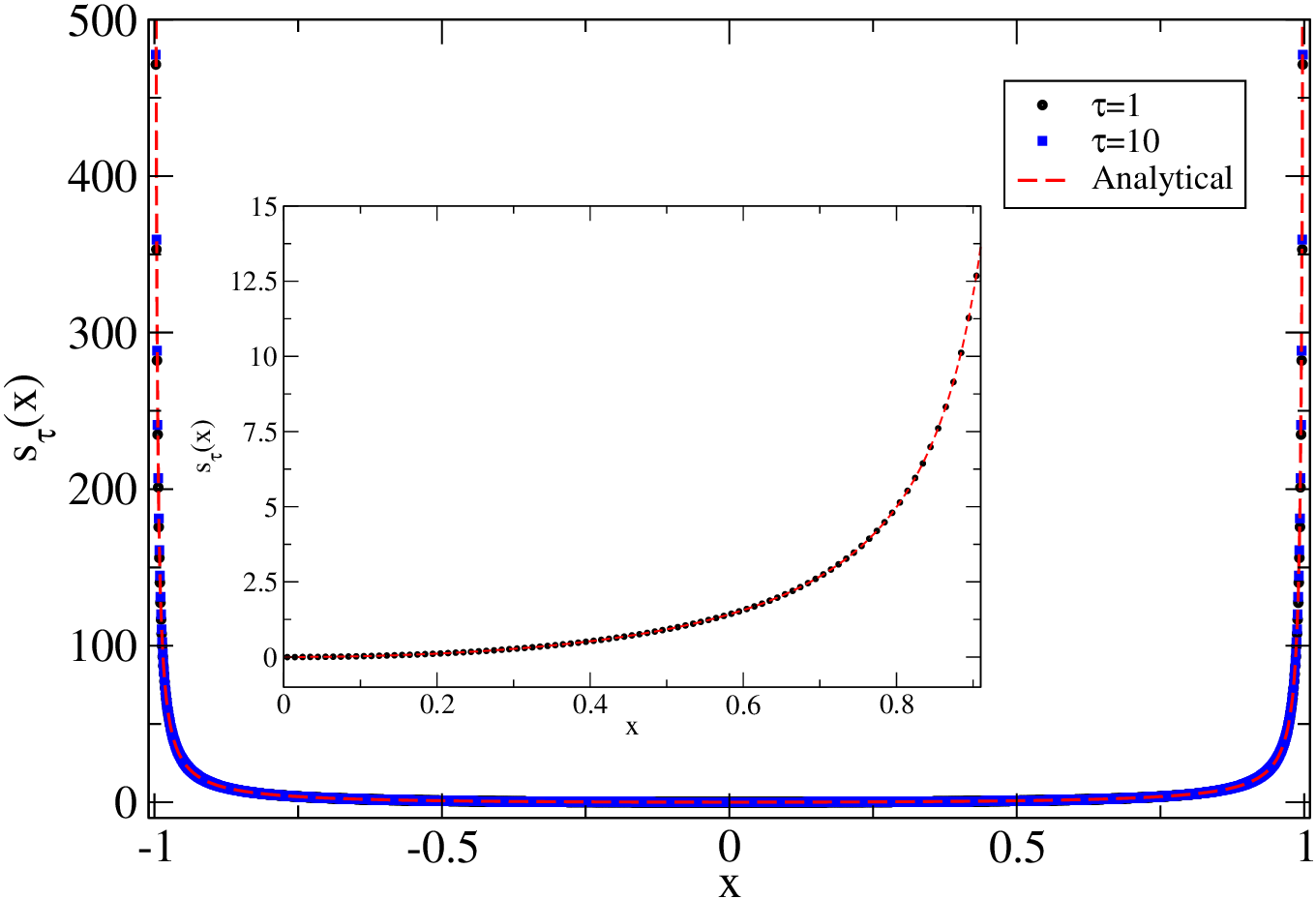}
\caption{The 
function, $s_{\tau}(x)$  (which is related to the SSD $P(x)$ via Eq.~\eqref{eq:ldf1}) for to an
  exponential distribution of running times,
  $g(t)=\exp(-t/\tau)/\tau$, for $\tau=1$ (black circles) and
  $\tau=10$ (blue squares). The red line shows the analytical solution
  (\ref{eq:ldf2}). The inset is a magnification of the central region
  $0<x<0.9$.}
\label{fig:ldfexp}
\end{figure}

The computational results for $s_{\tau}(x)$ are plotted in
Fig.~\ref{fig:ldfexp} for $\tau=1$ (black circles) and $\tau=10$ (blue
squares). The red line shows the analytical solution
(\ref{eq:ldf2}). The inset is a magnification of the central region
$0<x<0.9$, before the rapid increase in $s_{\tau}(x)$ near the edge of
the support. As can be clearly seen, the computational results for
both values of $\tau$ are almost indistinguishable, and the fit to the
analytical solution is nearly perfect.

\begin{figure}[t]
\includegraphics[width=0.98\linewidth,clip=]{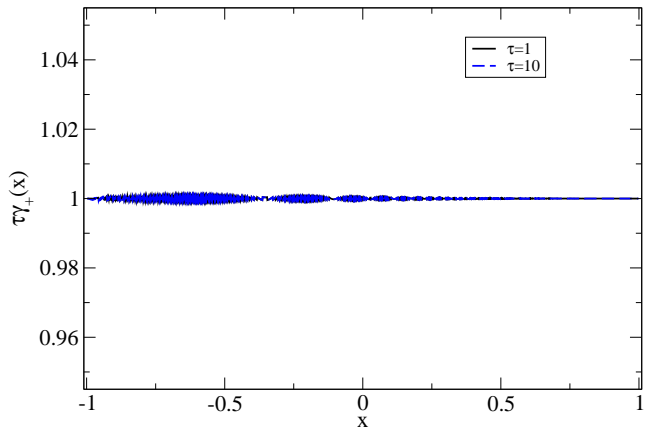}
\caption{The dimensionless local switching rate, $\tau\gamma_+(x)$ 
  corresponding to an exponential distribution of running times, for
  $\tau=1$ (black solid line) and $\tau=10$ (blue dashed line).}
\label{fig:gammaexp}
\end{figure}

In Fig.~\ref{fig:gammaexp} we plot the {\em dimensionless}\/ local
switching rate, $\tau\gamma_+(x)$, for $\tau=1$ and $\tau=10$. For
both values of $\tau$, the computational results exhibit excellent
agreement with the expected result that $\tau\gamma_+(x)=1$, up to
minor  (smaller than 0.1\%) discretization errors.  Notice
  the general trend that these errors almost disappear for $x>0.5$. This is expected since longer running times, which are less
  sensitive to space discretization, are required for right-moving
  particles to reach (and tumble) this region of the potential trap.

\subsection{Semi-Gaussian distribution}

\begin{figure}[ht]
\includegraphics[width=0.98\linewidth,clip=]{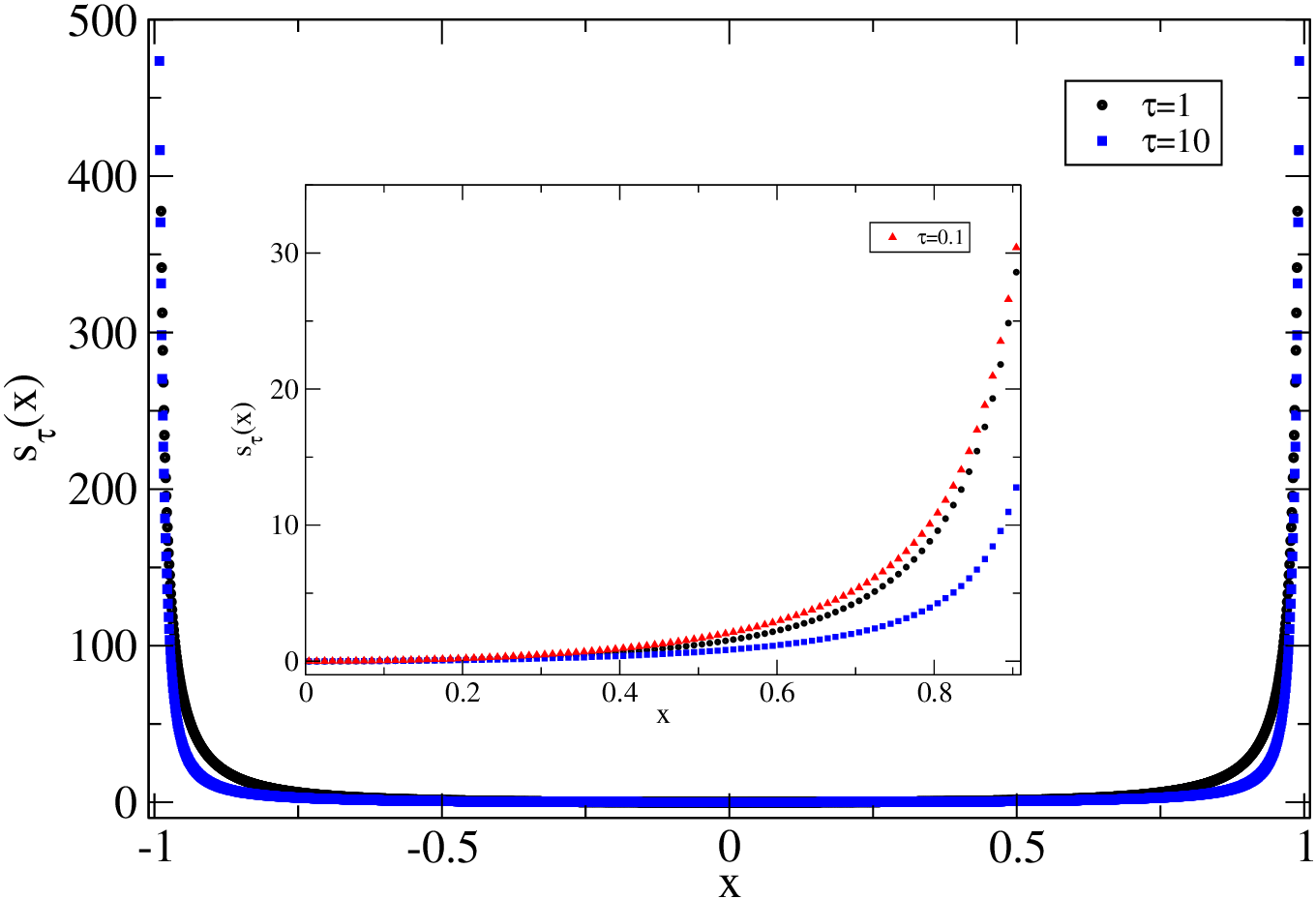}
\caption{The function $s_{\tau}(x)$ 
 (which is related to the SSD $P(x)$ via Eqs.~\eqref{eq:current3} and \eqref{eq:ldfdef})
corresponding to a
  semi-Gaussian distribution of running times [Eq.~(\ref{eq:gauss})]
  for $\tau=1$ (black circles) and $\tau=10$ (blue squares). The inset
  is a magnification of the central region $0<x<0.9$, with data also
  corresponding to $\tau=0.1$ which is plotted with red triangles.}
\label{fig:ldfgauss}
\end{figure}

Next, we consider the case where
\begin{equation}
  g(t)=\frac{2}{\pi\tau}\exp\left[-\frac{1}{\pi}
    \left(t/\tau\right)^2\right] \, .
  \label{eq:gauss}
\end{equation}
    Setting $\tau=1$ and $\tau=10$ as in section \ref{subsec:exp}, and defining
the function $s_{\tau}(x)$  as in Eq.~(\ref{eq:ldfdef}),
we present in Fig.~\ref{fig:ldfgauss} the results for
$s_{\tau}(x)$. 
The shape of the curves are quite similar to
the shape of the $s_{\tau}(x)$
in Fig.~\ref{fig:ldfexp}, i.e., grow moderately for $|x|\lesssim 0.9$
and exhibit a steep incline when approaching the edge of the
support. However, in the present case, the (exact) $s_{\tau}(x)$ does
depend on $\tau$. In the inset on Fig.~\ref{fig:ldfgauss}, we see a
magnification of the central region, including also results for
$\tau=0.1$ that are plotted with red triangles and deviate only weakly
from the data for $\tau=1$ (black circles). We note that for
$\tau=0.1$, the SSD becomes so small close to the edge of the support,
that it could not be computed beyond the range $|x|<0.95$ with double
precision floating point.

\begin{figure}[t]
\includegraphics[width=0.98\linewidth,clip=]{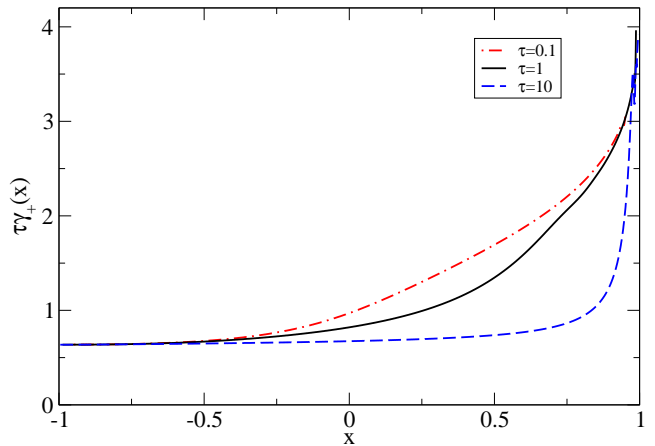}
\caption{The dimensionless local switching rate, $\tau\gamma_+(x)$
  corresponding to a semi-Gaussian distribution of running times, for
  $\tau=0.1$ (dashed-dotted red line), $\tau=1$ (black solid line),
  and $\tau=10$ (blue dashed line).}
\label{fig:gammagauss}
\end{figure}

\begin{figure}[b]
\includegraphics[width=0.98\linewidth,clip=]{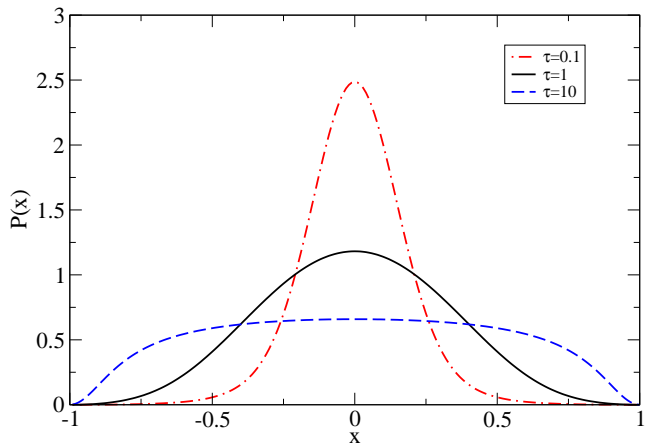}
\caption{The SSD, $P(x)$, corresponding to a half-t distribution of
  running times [Eq.~(\ref{eq:tdist})], for $\tau=0.1$ (dashed-dotted
  red line), $\tau=1$ (black solid line), and $\tau=10$ (blue dashed
  line).}
\label{fig:ldfalgeb}
\end{figure}

In Fig.~\ref{fig:gammagauss} we plot the dimensionless local switching
rate, $\tau\gamma_+(x)$, for $\tau=0.1$, $\tau=1$, and $\tau=10$. As
expected, the switching rate is not constant and, moreover, depends
also on $\tau$. We do, however, observe that all three curves are
monotonically increasing and approach the same values on the left and
right ends of the interval. These limits can be understood by noting
that the switching rate can be formally written as
\begin{equation}
  \gamma_+(x)=\left \langle \frac{g(t)}{1-G(t)}
  \right\rangle_{t=t_+(y\leq x,x)},
  \label{eq:gammalimits}
\end{equation} 
where the average, with the proper weights, is taken over all possible
trajectories leading directly to point $x$ from a point $y\leq x$, and
$t_+(y,x)$ is the travel time between the points [see
  Eq.~(\ref{eq:tplusab})]. Close to the left end of the support
($x\rightarrow -1$), most of the trajectories are short, and the
travel times $t_+(y,x)\ll\tau$; therefore
\begin{equation}
  \gamma_+(x\rightarrow -1)\rightarrow g(0).
  \label{eq:g0}
\end{equation}
In the present case this yields $\tau \gamma_+(x\rightarrow
-1)\rightarrow 2/\pi\simeq 0.637$, in excellent agreement with the
numerical data. On the right end of the support ($x\rightarrow 1$),
the different curves seem to approach the limit $\tau
\gamma_+(x\rightarrow 1)\rightarrow 4$. This limit can be related to
the asymptotic behavior of $s_{\tau}(x)$ near the edge of the support,
which becomes independent of $\tau$ because of the divergence of the
travel times in this region (i.e., the diminishing of the
velocity). Explicitly, for $x\rightarrow 1$, Eq.~(\ref{eq:current1})
is well approximated by
\begin{equation}
  \frac{d\ln [I(x)]}{dx}\simeq- \frac{\gamma_+(x)}
       {v_+(x)}=\frac{\mu \gamma_+(x)}{\sigma_0+F(x)}.
  \label{eq:currentapp1}
\end{equation}
For the force function (\ref{eq:force}) with $\sigma_0=1$, we obtain
from Eqs.~(\ref{eq:ldf1}) and (\ref{eq:currentapp1}) that
\begin{eqnarray}
  \frac{s_{\tau}(x\rightarrow 1)}{\tau}&\simeq&
  \frac{2\gamma_+(x\rightarrow 1)\mu}
       {\pi}
  \tan\left[\frac{\pi}{4}
    \left(1+x\right)\right]\nonumber \\
  &\simeq&\frac{8\gamma_+(x\rightarrow 1)\mu}{\pi^2}\frac{1}{1-x}.
  \label{eq:ldfapp1}
\end{eqnarray}
This result applies to any function $g(t)$ (for the force function
considered herein), as long as $\gamma_+(x)$ does not vanish when
$x\rightarrow 1$ (as in the example discussed in the following
subsection).  In particular, it implies that $\tau \gamma_+(x\to1)$ is
independent of $\tau$, in agreement with our numerical results 
  and consistent with the large-deviation principle.  It is easy to
check that the LDF (\ref{eq:ldf2}) of the exponential distribution
function [in which case $\gamma_+(x)=1/\tau$], has the same asymptotic
form as the general Eq.~(\ref{eq:ldfapp1}).

\subsection{Half-t distribution}

\begin{figure}[t]
\includegraphics[width=0.98\linewidth,clip=]{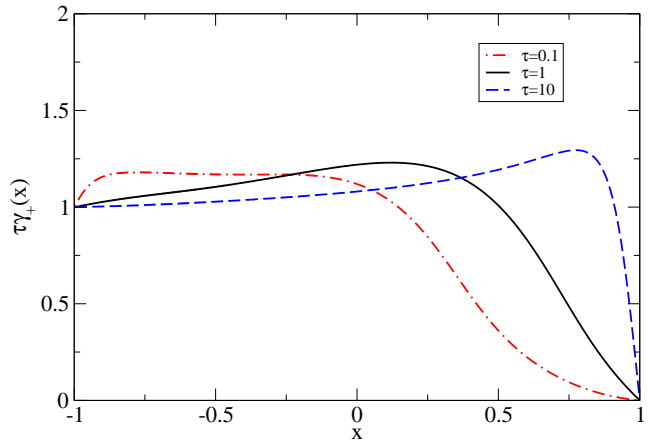}
\caption{The dimensionless local switching rate, $\tau\gamma_+(x)$
  corresponding to a half-t distribution of running times, for
  $\tau=0.1$ (dashed-dotted red line), $\tau=1$ (black solid line),
  and $\tau=10$ (blue dashed line).}
\label{fig:gammaalgeb}
\end{figure}

Consider the distribution function of running times 
\begin{equation}
  g(t)=\frac{1}{\tau}\left[1+(t/\tau)^2\right]^{-3/2},
  \label{eq:tdist}
\end{equation}
which is a special case of the algebraically decaying {\em half-t
  distribution}\/ with average running time equal to $\tau$ and
diverging higher moments. Fig.~\ref{fig:ldfalgeb} shows the SSDs (which
have been normalized to unity) for $\tau=0.1$, $\tau=1$, and
$\tau=10$. Since the distribution function $g(t)$ is fat-tailed,
it is not surprising to find that the SSD is rather flat for
$\tau=10$. For $\tau=1$ and $\tau=0.1$, the SSD adopts a bell-shape,
but one which is far less narrowly peaked compared to the SSDs corresponding to
the exponential distribution (subsection \ref{subsec:exp}) with
similar $\tau$. This is the reason why, in this case, we do not plot
the functions $s_{\tau}(x)$. 
In fact, it is not clear whether the large-deviation
principle applies here, as can be understood from
Fig.~\ref{fig:gammaalgeb} where we plot the dimensionless switching
rate for the different values of $\tau$. All three curves approach the
limit $\tau\gamma_+(x)\rightarrow 1$ for $x\rightarrow -1$, which is
consistent with Eq.~(\ref{eq:g0}). On the other end of the interval,
we see that all three curves converge to the limit
$\tau\gamma_+(x\rightarrow 1)\rightarrow 0$. As noted above, the
general asymptotic form (\ref{eq:ldfapp1}) does not hold in such
cases. The vanishing of $\gamma_+(x)$ close to the right edge of the
support means that the stopping probability of the particle does {\em
  not} assume a local exponential form, see Eq.~(\ref{eq:comgamma1}),
which is required for the SSD to take the form $P(x)\sim
\exp[-s(x)/\tau]$ in the large-deviation regime.

\section{Discussion}
\label{sec:disc}

To summarize, we have developed a numerical method for computing the
SSD of a non-Markovian RTP, characterized by a waiting time
distribution $g(t)$ between tumbling events, and trapped by an
external potential in 1D. Our method succeeds in obtaining numerical results
even in regions of space for which the SSD is extremely small, i.e.,
representing very rare events. These events are associated with
situations in which the RTP is close to the edges of the SSD's
support, and they become rarer as the mean time between tumbles $\tau$
is decreased.  Our method does not involve generating (naive or
biased) realizations of the RTP's dynamics. Instead, it is based on an
analytical derivation that leads to a set of integral equations whose
solution yields the SSD. We essentially solve these integral equations
numerically.

To illustrate our method, we applied it to three different
distributions $g(t)$: (i) Exponential,  (ii) semi-Gaussian, and (iii)
  half-t.
In the first example, one recovers the
standard (Markovian) RTP, whose SSD is known exactly. This case serves
as a useful benchmark for our method.
The last example leads to very different
scaling behaviors since $g(t)$ is fat tailed.

The computation of the SSD and the function $s_{\tau}(x)$ that is
closely related to it, also enables us to compute the LDF,
$s(x)$.  This is done by combining results for different values of
  $\tau$,  and considering the asymptotic behavior of
    $s_{\tau}(x)$ in the limit $\tau\rightarrow 0$.
 Near the edge of the SSD's support, the LDF is characterized by
  (i) a diverging function, and (ii) the asymptotic value of the
  position-dependent tumbling rate $\gamma_+(x)$. These depend on the
  form of the external force, $F(x)$, and on the distribution function
  of the waiting times, $g(t)$ [under mild assumptions regarding
    $g(t)$], but not on $\tau$. For the specific external force
(\ref{eq:force}), we obtained analytically
that $s(x)$ diverges as a power law. This particular
behavior appears to be related to a special property of the force
(\ref{eq:force}): Its derivative, $F'(x)$, vanishes at the edge. In the
more generic case in which this does not occur, other behaviors are to
be expected (e.g., logarithmic divergence
\cite{SmithFarago22}).  Note that $s(x)$ is an extremely useful
object.  Given $s(x)$ for a specific potential,
one can immediately compute from it the rate function, 
  $\Phi(z)$ (see definition in~\cite{SmithFarago22, Smith2023}) that
describes dynamical fluctuations of the RTP's position in the absence
of external forces.  From the knowledge of $\Phi(z)$, one can
  then work in the opposite route and obtain $s(x)$ for other
  potentials~\cite{Smith2023}.

Our method (perhaps with minor adjustments) may be applicable to
non-confining potentials as well, and could therefore prove useful to
study escape or first-passage statistics, or behavior within periodic
potentials, for non-Markovian RTPs. It would be also interesting to
extend our method to more general settings, e.g., to higher dimensions
or to other possible non-Markovian active models (e.g., one could
consider non-Markovian versions of the active Brownian
particle).  Finally, it is also left to check how well the
  numerical method works in some ``exotic'' cases like the Pearson
  random walk model~\cite{prw} in 1D, where the distribution function
  $g(t)$ of tumbling times is not continuous but discrete.

\subsection*{Acknowledgments}

We acknowledge support from the Israel Science Foundation (ISF)
through Grants No.~1258/22 (OF) and 2651/23 (NRS).






\begin{thebibliography} {99}
\bibitem{footnote1} Note that often in the literature the standard RTP
  is referred to as a non-Markovian model too, the interpretation
  being that its future dynamics do not depend on its instantaneous
  position alone.
\end{thebibliography}
\end{document}